\documentclass[twocolumn,prd,nofootinbib,aps,prl,floats,floatfix,amsmath,amssymb,longbibliography,secnumarabic]{revtex4-1} %
\usepackage[english]{babel}

\usepackage[letterpaper,top=2cm,bottom=2cm,left=3cm,right=3cm,marginparwidth=1.75cm]{geometry}

\usepackage{amsmath}
\usepackage{graphicx}
\usepackage[colorlinks=true,citecolor=red]{hyperref}

\newcommand{\be}{\begin{equation}}
\newcommand{\ee}{\end{equation}}

\newcommand{\bea}{\begin{eqnarray}}
\newcommand{\eea}{\end{eqnarray}}
\newcommand{\nn}{\nonumber}

\begin{document}

\title{Quantum coherence between mass eigenstates of a neutrino cannot be destroyed by its mass-momentum entanglement}
\author{James M.\ Cline}
\email{jcline@physics.mcgill.ca}
\affiliation{McGill University Department of Physics \& Trottier Space Institute, 3600 Rue University, Montr\'eal, QC, H3A 2T8, Canada}
\affiliation{Niels Bohr International Academy,The Niels Bohr Institute,
Blegdamsvej 17, DK-2100 Copenhagen Ø, Denmark }

\begin{abstract}
In the latest of a series of papers (\url{https://arxiv.org/pdf/2410.21850}), it has been claimed that neutrinos cannot oscillate as a consequence of their mass
mixing.  I clarify that this arises from the unrealistic assumption that neutrinos will always be in nearly exact eigenstates of energy, and ignoring the spatial dependence of the neutrino wave function.  Taking into account the latter, neutrinos will be observed to oscillate in precisely the expected manner.

\end{abstract}

\maketitle

Recently Ref.\ \cite{Zheng:2024bxr} reiterated the claim that neutrinos cannot oscillate as a consequence of mixing and mass-squared differences.  The essence of the claim is that a flavor state emitted from a weak interaction will be in a nearly exact eigenstate of energy.  In this case, the state evolves by an overall phase $e^{-iEt}$, and therefore there are no oscillations between the mass eigenstates appearing in the superposition of states.  However, even if there are no oscillations in time, such a state has oscillations in space, due to the momentum differences.  Suppose the neutrino is moving along the $z$ direction.
Then the wave function for a $\nu_e$ state is
\bea
    \psi_e(t,z) &=& e^{-iEt}\sum_i U_{ei} e^{i p_i z}\\
        &\cong & e^{-iE(t-z)}\sum_i U_{ei} e^{-iz m^2_i/2E}\nn\,.
\eea
Despite the probability distribution being stationary in time, due to the spatial oscillations, an observer at a distance $L$ from the production site will measure a transition probability that exactly agrees with the usual prediction, conventionally based on equal momenta and different energies, averaged over the oscillation time.  The observer sees only the stationary oscillation pattern, not whether the interference occurs in time or in space.  There is no reason why it should be purely one or the other; this would require a very careful preparation of the initial state.  In general there will be uncertainties in both the energies and momenta of the interfering states.  Fortunately, it does not matter for observations.
Issues involving decoherence and wave packets have already been carefully considered in the literature, for example, Refs.\ \cite{Giunti:1997wq,Giunti:2006fr}.

The author of \cite{Zheng:2024bxr} proposes that oscillations are caused by virtual $Z$ boson exchanges, due to flavor off-diagonal couplings $U_{ij}$.  
(Here $U$ is no longer the PMNS matrix, but rather some unmotivated, hypothetical mixing matrix in flavor space.)
This ignores the fact that in the Standard Model of particle physics, the $Z$ boson couples diagonally to neutrino flavors, {\it e.g.,} section 10.1 of Ref.\ \cite{ParticleDataGroup:2024cfk}.  Were it not the case, a two-loop diagram involving $W$ and $Z$ exchange would induce the rare decay $\mu\to e\gamma$
at the level of $10^{-7\,}|U_{\mu e}|^2$, requiring $|U_{\mu e}|\lesssim 10^{-3}$, much smaller than the typical size of mixing in the PMNS matrix.

I thank Kimmo Kainulainen for valuable discussions.

\bibliographystyle{utphys}
\bibliography{sample}

\end{document}